# Diving into Buoyancy

## Exploring the Archimedes Principle Through Engineering

BY MOUHAMADOU THIAM

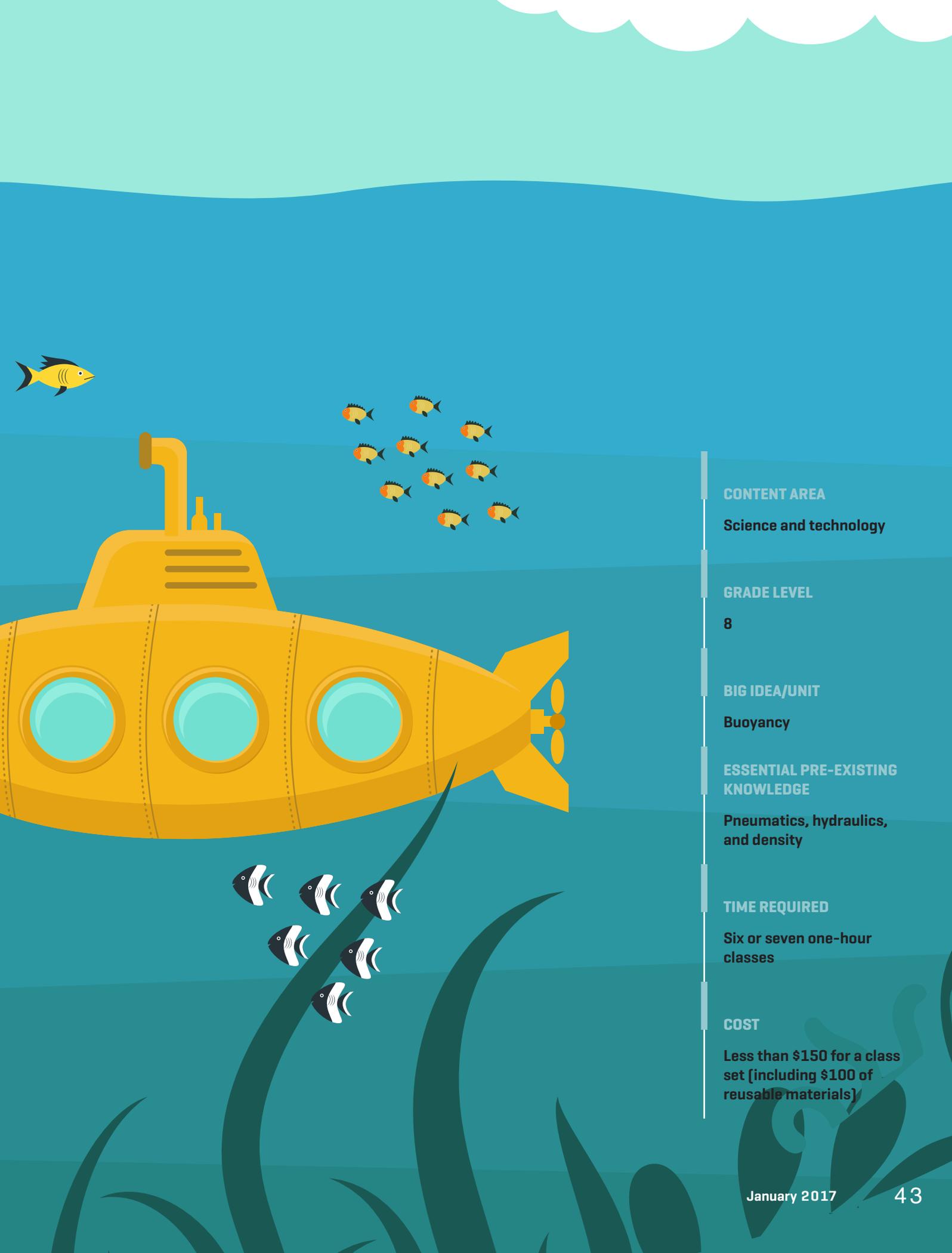

**CONTENT AREA**
Science and technology

**GRADE LEVEL**
8

**BIG IDEA/UNIT**
Buoyancy

**ESSENTIAL PRE-EXISTING KNOWLEDGE**
Pneumatics, hydraulics, and density

**TIME REQUIRED**
Six or seven one-hour classes

**COST**
Less than $150 for a class set (including $100 of reusable materials)

In our daily lives, we observe objects sinking, floating, or rising when immersed in a fluid. The *Archimedes principle*, which explains an object's behavior when immersed in a fluid, is important in fluid mechanics; however, it is a relatively complex concept for middle school students to grasp, as they often harbor misconceptions. To initiate conceptual change among students regarding the misconception "heavy objects sink and light objects float," I created a project during which students build a stable submarine that uses fluid transfers to move up, down, and forward while carrying a load. Students must take into account several variables, from the design of the submarine to the choice of materials. Additionally, students write a report that includes a user manual, challenges they encountered and how they overcame those challenges, and a detailed text that links theory to their submarine.

## Context

During my first two years of teaching, students participating in this project were allowed to build any device that used fluid transfers. Students usually limited their choices to hydraulic devices such as a dentist's chair, an elevator, or a parking lot. I found these items to be too simple to fully demonstrate a proficient understanding of fluid transfers. Therefore, I designed this complex engineering project, which gives students the opportunity to build a submarine. Middle level students often hold several misconceptions regarding the Archimedes principle. One of them, mentioned by Thouin (2004), is that a heavier object will tend to sink, whereas a lighter one will float. Through this engineering project, students learn how to modify the density of an object to make it float or sink. Contrary to some engineering projects where students receive a list of materials and are guided with steps to the same final product, I put them in a real-life situation in which they had to design, build, and test their own prototypes.

## Problem

Fort McMurray, Alberta, is the capital city of the oil sands, with one of the largest crude oil reserves in the world. Most of the equipment needed to extract crude oil must travel from the south (far outside of the city) to the north of the province to reach the extraction sites. These loads disrupt traffic on Highway 63, already considered one of the most dangerous highways in Canada. However, the Athabasca River, with a length of 1,300 km (808 mi.), flows northeast across Alberta (RAMP 2015) and could be used to carry equipment to sites. To interest my students in this project, a performance task (Figure 1) is presented to them. Students and their families frequently use Highway 63 to reach southern destinations and therefore are relatively aware of the numerous fatalities that occur on this road. Students also tend to notice the highway's abundance of commercial vehicles, which travel between north and south, and take into consideration the vehicles' gas emissions.

## Procedure

This project is designed for grade 8 students; however, it can be introduced at any grade level where the Archimedes principle is covered. This project was completed during six one-hour blocks.

### Block 1

The performance task and rubric (Figure 2) were

**FIGURE 1: Performance task: Build a submarine**

> *You are a mechanical engineer. Roads in Fort McMurray are in poor condition due to the extensive use of trucks. In addition, major traffic jams are common. A public contract from one of the major oil companies invites engineers to build a stable submarine, which should be capable of moving through the Athabasca River to transport materials. Because the Athabasca River has different depths, the submarine must be able to move up and down in the water through fluid transfer. Because the company is sensitive to the environment, your submarine should predominantly be made of recycled or reclaimed materials. You will prepare a report that includes your submarine's construction stages and information about materials used and how you applied fluid properties and materials' behavior to the project.*

FIGURE 2: Rubric used to assess students

| Criteria | 1 | 3 | 5 |
|---|---|---|---|
| Materials | Low use of recycled or reclaimed materials. | Partial use of recycled or reclaimed materials. | Predominant use of recycled or reclaimed materials. |
| Instructions | Few detailed pictures/drawings. Unclear and incomplete instructions on the use of the submarine. | Some detailed pictures/drawings. Sometimes clear and detailed instructions on the use of the submarine. | Detailed pictures/drawings. Clear and detailed instructions on the use of the submarine. |
| Journal | Few details about challenges and the methods used to overcome them. | Some details about challenges and the methods used to overcome them. | Describes several challenges and the methods used to overcome them. |
| Understanding | Provides limited informative text to explain fluid properties and the behavior of materials immersed in it. | Provides general informative text to explain fluid properties and the behavior of materials immersed in it. | Provides informative text to explain deeply and thoroughly fluid properties and the behavior of materials immersed in it. |
| Impact (×2) | The submarine has limited capability to carry a load and is unstable. | The submarine is generally capable of carrying a load and is sometimes stable. | The submarine can carry a load and is always stable. |
| Total | /30 | | |

presented to students. Depending on the group and whether time allows, the teacher may work with students to create a common rubric for the project. I also showed students a large, long plastic container (42.7 × 24.4 × 18 in.) that was three-quarters filled with water and explained that they would test their submarines in it for final evaluation. Give students an opportunity to ask questions about the task and rubric. My students most frequently ask about the tasks that the submarine must perform; teachers should make sure that students understand that the submarine must carry a load (two large spoons, in this case) while performing tasks such as moving upward, downward, and forward. Emphasize that a report is due at the end of the project, which will account for half of the points for the overall project.

Students are then informed that they can work in teams or independently; however, they are highly encouraged to work in teams of up to three members because the project requires collaboration and can be overwhelming. Five minutes are allocated so students can form groups. After teams are formed, students start brainstorming ideas for their submarine design. The teacher circulates, answering students' questions and asking questions of groups to make sure students are aware of the rubric's criteria: "What is your estimate of the percentage of recycled and reclaimed materials used in your device? What tasks are you expecting your device to perform? What components must be present in your report? What is the role of each member of your team and how were those roles assigned?"

### Block 2

This block is exclusively reserved for the submarine design, which is crucial because students have to work with several variables, such as the submarine's dimensions, the types of materials, and the fluid transfer mechanical systems. Hydraulic and pneumatic systems are introduced to students before the project. For this project, I decided not to show students previous years' models to ensure a variety of final products. Students most frequently ask me what types of materials they should use to build their submarine. The teacher's role is to guide students through the design process by prompting them with questions such as "How should the materials used in the construction of the submarine behave when immersed in water? What properties of the materials should be taken into account?" Students are invited to show the teacher their blueprints and explain how their submarine would perform the necessary tasks while transporting the two large spoons. To prepare students for creating their blueprints, I ask them to conduct online research in small groups for a period of 15 minutes. Students explore how a particular fish moves upward and dives into the water. They then discuss similarities between the swim bladder in a fish and the ballast of a submarine. The teacher's role at this point is to give feedback on what students could do to improve their designs. Questions posed by the teacher might include:

- Which parts of the submarine will make your submarine move forward, up, and down?

- What fluids will you use and how will they be transferred so your submarine performs the movements?

- Why do you think your submarine will be able to transport two spoons and remain stable?

- How will your choice of material impact the performance of your submarine?

Prior to experimentation, the teacher approves the blueprints. After receiving feedback, students continue to work on their designs. At the end of the block, none of the designs will look similar. Students discuss amongst themselves the materials that each member of the group should bring so they can start building their prototype. Teachers should have available some syringes (60 mL), clear plastic tubing compatible with the syringes, tubing connectors, straws, glue, and balloons in case students forget to bring them, because most prototypes will use these materials. Syringes, clear plastic tubing, and tubing connectors can be bought at most science lab suppliers. Usually, when needed, students can continue working on their designs during the next block. Each team is assigned a drawer in the classroom to store the materials between classes.

### Blocks 3 to 5

Teachers should discuss and emphasize the crucial aspect of safety in the science classroom before students begin building and testing their prototypes. Because students will be using sharp tools to cut certain materials, every student must wear safety glasses to protect their eyes in case the blade breaks or a piece of material flies. Students are reminded to be careful when using cutting tools.

Students' submarines are usually good at sinking and rising. The challenging part is moving them forward while keeping them stable. Stability is therefore the most challenging aspect of the project. Students usually think that having only one bottle in their design will make a stable submarine because they attempt to test their submarine without the load. Then, after placing the load, they notice that the submarine tilts as it is sinking or rising. As students will need to run several tests, the role of the teacher is to give them feedback by questioning them regarding the reasons why the submarine failed to perform a certain task. Teachers can ask, "From which side did the submarine tilt? What is the reason for this imbalance? How could the system achieve balance?"

To rise or sink the submarine, my students commonly used balloons inside bottles. Teams who used two bottles with a balloon inside each encountered some challenges. When air is not blown into the balloons simultaneously, the submarine tilts. I ask students, "How does the system responsible for the rise of your submarine work? What observations do you make when you fill the balloons with air? What impact does this inflating time gap between the two balloons have on the stability of the submarine? What changes should be made to simultaneously inflate the balloons? Students are excited before the first tests but some frustration may arise by the tenth test, which will be needed as they continue to run new tests with modifications to the original design. Students create designs of their submarine in their science notebooks. Information on the modifications done on the submarine and its new behavior are also stored in their science notebook. Additionally, students note for each trial how long their submarine was able to float and whether it was able to move up, down, and forward. To help students overcome their frustration after several trials, they are asked to get their blueprints approved by the teacher after any modification.

My students usually create three different methods for moving the submarine forward. The most common method involves a tube at the back of the submarine, into which they blow air to force the submarine to move forward. The second method uses a propeller, which usually works when the submarine is not loaded, because the propeller adds weight. For this method, I ask students, "What properties can explain the fact that the submarine no longer advances when you add the load? What characteristics of the propeller or the submarine must be modified to increase the efficiency of the propulsion?" The third method involves the use of effervescence pills, which produce bubbles when they come into contact with water and propel the submarine. Some student models are shown in Figure 3.

### Block 6

As they build and modify their submarines, students must keep records of all their steps to avoid any duplicated trials throughout the process. This is done as students are building and modifying their submarines. However, an additional block is needed to process all information, including the challenges students encountered and how these were overcome.

My students conduct final testing over two blocks, but this amount of time may vary depending on the number of teams. Each team is allowed three trials, with the final score based on the best trial. Students are scored on whether their submarine is stable for at least one minute and able to move up, down, and forward. The report and the submarine performance were given a summative assessment using the rubric (Figure 2).

### Assessment

Students are formatively assessed throughout the project. Students' understandings are formatively assessed at the beginning of the unit and at the end of the project using four formative assessment probes (Comparing Cubes, Floating Logs, Floating High and Low, and Solids and Holes) from Keeley, Eberle and Tugel (2007). Each formative assessment probe begins with a scenario; students choose the best answer(s) from a given list of statements and ultimately provide an explanation of their thinking. My students' responses to these probes attested that they had gained a better understanding of density after completing the project compared to the beginning of the unit. Here are some questions that I ask students to help them better understand density:

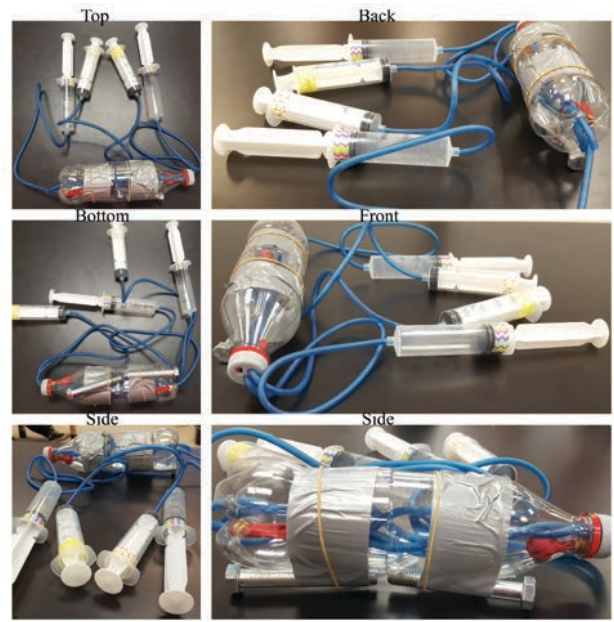

**FIGURE 3:** Samples of student submarines

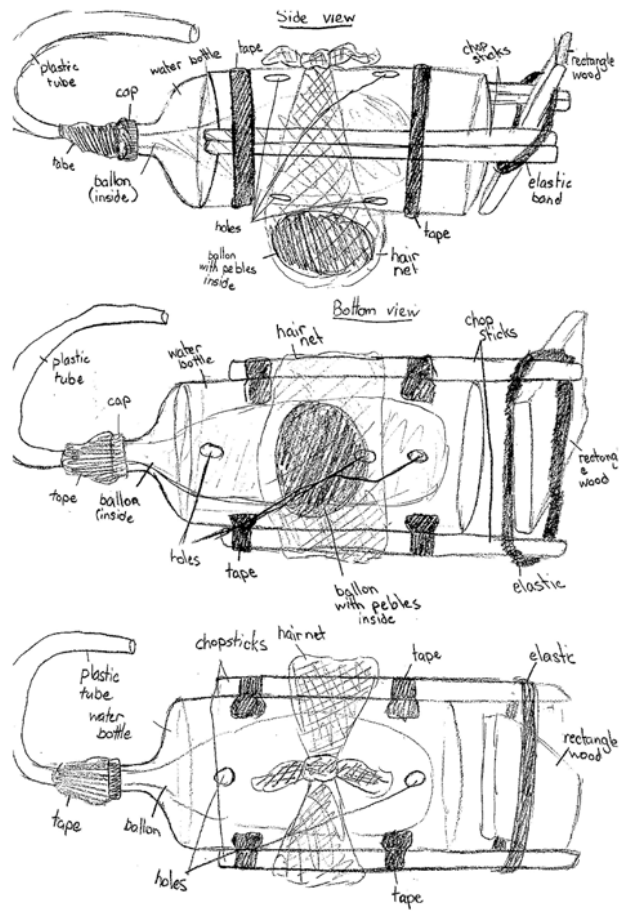

# Connecting to the *Next Generation Science Standards* (NGSS Lead States 2013)

- The chart below makes one set of connections between the instruction outlined in this article and the *NGSS*. Other valid connections are likely; however, space restrictions prevent us from listing all possibilities.
- The materials, lessons, and activities outlined in the article are just one step toward reaching the performance expectations listed below.

### Standards

MS-PS2-2 Motion and Stability: Forces and Interactions
*www.nextgenscience.org/pe/ms-ps2-2-motion-and-stability-forces-and-interactions*

MS-ETS1: Engineering Design
*www.nextgenscience.org/dci-arrangement/ms-ets1-engineering-design*

### Performance Expectations

MS-PS2-2. Plan an investigation to provide evidence that the change in an object's motion depends on the sum of the forces on the object and the mass of the object.

MS-ETS1-3. Analyze data from tests to determine similarities and differences among several design solutions to identify the best characteristics of each that can be combined into a new solution to better meet the criteria for success.

MS-ETS1-4. Develop a model to generate data for iterative testing and modification of a proposed object, tool, or process such that an optimal design can be achieved.

| DIMENSIONS | CLASSROOM CONNECTIONS |
|---|---|
| **Science and Engineering Practice** | |
| Developing and Using Models | Students build a stable submarine that uses fluid transfers to move up, down, and forward, while carrying a load, to solve a community issue. |
| **Disciplinary Core Idea** | |
| MS-PS2-2<br>• The motion of an object is determined by the sum of the forces acting on it; if the total force on the object is not zero, its motion will change. The greater the mass of the object, the greater the force needed to achieve the same change in motion. For any given object, a larger force causes a larger change in motion. [MS-PS2-2] | Students answer questions that require knowledge of density. Students write a user manual that links a theory to their submarine. |
| ETS1.C: Optimizing the Design Solution<br>• The iterative process of testing the most promising solutions and modifying what is proposed on the basis of the test results leads to greater refinement and ultimately to an optimal solution [MS-ETS1-4] | Students build, test, and evaluate their submarine and make iterative changes to it in order to achieve maximum efficiency. |
| **Crosscutting Concept** | |
| Cause and Effect | Students manipulate their submarine model to answer the following questions:<br>• What do you observe when you fill the balloons with air?<br>• What impact does this time gap (caused by inflating) have on the stability of the submarine?<br>• What changes should be made to simultaneously inflate the balloons? |

- A solid ball is placed in a container filled with water. Half of the ball floats above the water's surface and the other half is in the water. What can you do to cause a larger part of the ball to be under the water?

- A log was cut from a tree and put in water. The log floats on its side, with half of it above the water surface. Another log was cut from the same tree. This log was twice the size of the other and twice as wide. How does the larger log float compared to the smaller?

- A solid and thin object, when placed in water, floats. If you poke holes in this object, what observations would you make when it is put back into the water?

Students receive feedback throughout the activity to improve their submarine. In addition to building a submarine, a student's report is also important because of the amount of detailed information required. To guide students through the writing of their submarine's report, a set of directions (see Online Supplemental Materials) is given to them. The report is important because it allows students to create a document similar to the ones engineers use to guide others through their choice of the materials, the steps followed to overcome challenges, the user's manual for the device, and how theory is applied to build a useful product. A checklist (see Online Supplemental Materials) created from the rubric can be handed out to students to help them meet all required elements in the rubric. Teams are highly encouraged to electronically share their report with the teacher to receive feedback.

## Conclusion

This project, although complex, has been welcomed by my students, according to feedback I have received. The point that came up most often is that they had fun building a device on their own and testing it: We do not often have the opportunity to build devices in science classes. Students tended to reflect more negatively on the report, which they considered of little importance given that the main objective, in their opinion, was building a submarine. To overcome this, I now differentiate the report portion by allowing students to choose how they will present their reports (e.g., multimedia, oral, or written presentations). Even if more time is necessary, this option will make the report more engaging to students.

Through formative assessments, my students showed a better understanding of the concept of density. Additionally, they used the terms "more dense" and "less dense" instead of "heavy" and "light" in their reports when explaining the behavior of their submarines. Archimedes' principle and concepts of mass, volume, and density were covered throughout the unit. Because the formative assessments administered to students were carried out before the project (at the beginning of the unit) and at the end of the project, it becomes difficult to measure with certainty the effect of this project on conceptual change in students. Therefore, I suggest that these pre- and posttests be administered just before and right after the project to measure its impact on students' misconceptions. ●

### ONLINE SUPPLEMENTAL MATERIALS

Student report instructions—*www.nsta.org/scope1701*

**Mouhamadou Thiam** (*mouhamadou.thiam@nbed.nb.ca*) is a teacher at Riverview East School in New Brunswick, Canada.